\documentclass[aps,prd,superscriptaddress,twocolumn]{revtex4-1}

\usepackage{amsfonts}
\usepackage{amsmath}
\usepackage{graphicx}
\usepackage{xcolor}
\usepackage{verbatim}
\usepackage[dvips]{epsfig}
\usepackage{float}
\usepackage{epsfig,subfig}
\usepackage{epstopdf}
\usepackage{hyperref}
\usepackage{cleveref}
\usepackage{placeins}

\begin{document}

\title{Quantum phase transition of two-level atoms interacting with a finite radiation field}

\author{L. F. Quezada}
\email{lfqm1987@ciencias.unam.mx}
\affiliation{Instituto de Ciencias Nucleares, Universidad Nacional Aut\'{o}noma de M\'{e}xico, 04510 Ciudad de M\'{e}xico, M\'{e}xico}
\affiliation{Centro de Ciencias de la Complejidad, Universidad Nacional Aut\'{o}noma de M\'{e}xico, 04510 Ciudad de M\'{e}xico, M\'{e}xico}

\author{A. Mart\'{i}n-Ruiz}
\email{alberto.martin@nucleares.unam.mx}
\affiliation{Instituto de Ciencias Nucleares, Universidad Nacional Aut\'{o}noma de M\'{e}xico, 04510 Ciudad de M\'{e}xico, M\'{e}xico}
\affiliation{Centro de Ciencias de la Complejidad, Universidad Nacional Aut\'{o}noma de M\'{e}xico, 04510 Ciudad de M\'{e}xico, M\'{e}xico}

\author{A. Frank}
\email{frank@nucleares.unam.mx}
\affiliation{Instituto de Ciencias Nucleares, Universidad Nacional Aut\'{o}noma de M\'{e}xico, 04510 Ciudad de M\'{e}xico, M\'{e}xico}
\affiliation{Centro de Ciencias de la Complejidad, Universidad Nacional Aut\'{o}noma de M\'{e}xico, 04510 Ciudad de M\'{e}xico, M\'{e}xico}
\affiliation{El Colegio Nacional, Ciudad de M\'{e}xico, M\'{e}xico}

\begin{abstract}
We introduce a group-theoretical extension of the Dicke model which describes an ensemble of two-level atoms interacting with a finite radiation field. The latter is described by a spin model whose main feature is that it possesses a maximum number of excitations. The approach adopted here leads to a nonlinear extension of the Dicke model that takes into account both an intensity dependent coupling between the atoms and the radiation field, and an additional nonlinear Kerr-like or P\"{o}sch-Teller-like oscillator term, depending on the degree of nonlinearity. We use the energy surface minimization method to demonstrate that the extended Dicke model exhibits a quantum phase transition, and we analyze its dependence upon the maximum number of excitations of the model. Our analysis is carried out via three methods: through mean-field analysis (i.e. by using the tensor product of coherent states), by using parity-preserving symmetry-adapted states (using the critical values obtained in the mean-field analysis and numerically minimizing the energy surface) and by means of the exact quantum solution (i.e. by numerically diagonalizing the Hamiltonian). Possible connections with the $qp$-deformed algebras are also discussed. 
\end{abstract}

\maketitle

\section{\label{sec:level1} Introduction}

The Dicke model (DM) \cite{Dicke} is one of the simplest quantum systems describing a single bosonic mode interacting collectively with a set of $N$ two-level systems (e.g. a cold two-level atomic cloud). Due to its importance in laser physics and quantum optics, the DM has been extensively studied both analytically and experimentally during the last decades. An important feature of this model is the presence of a phase transition between the normal and the super-radiant behavior \cite{Hepp-Lieb, Hepp-Lieb2, Wang-Hioe}. This interesting quantum phase transition (QPT) has been experimentally observed in a super-fluid gas in an optical cavity \cite{Baumann, Nagy}, although this development was achieved using time-dependent fields dressing the system. Furthermore, in similar models with the atoms in a wave guide, an specific super-radiance of just one photon has recently been investigated due to its potential applications in high-speed quantum communications \cite{Chen1,Chen2,Chen3,Chen4}.

Since the DM is a standard model in quantum optics, its extension in different directions is of general interest. It has been generalized in different ways, e.g. when the interaction between the radiation field is no longer linear in the field variables, i.e. allowing an intensity-dependent coupling. In a similar vein, there are models incorporating nonlinear terms in the photon number operator in order to investigate the evolution of the hybrid in the presence of a Kerr-like medium. In this framework, the application of the quantum algebras has proved useful in obtaining exactly solvable models in quantum optics \cite{Dobrev}. For example, many authors have studied the $qp$-deformed Jaynes-Cummings (JC) and Dicke models, which are obtained by replacing the usual creation and annihilation bosonic operators by their corresponding $qp$-deformed operators \cite{Chaichian, Bonatsos, Crnugelj, Recamier, SantosSnchez}. Interestingly, the various versions of the JC model can be described in a unified formalism based on the generalized deformed oscillator algebra, in which different choices of the deformation function provides the description of different physical systems, as summarized in Table 1 of Ref. \onlinecite{Bonatsos}. Despite their mathematical richness, these models mutilate the basic algebraic structure of the atom-field hybrid. 

In the light of the aforementioned studies, in this paper we introduce a group-theoretical extension of the Dicke model which is based on the fact that the Heisenberg-Weyl algebra HW$(1)$ describing the usual bosonic mode can be obtained by contraction of the SU$(2)$ algebra \cite{Ricci}. In short, to describe the radiation field, we propose a spin model which exhibits the main properties of the Kerr medium. The model is formulated in terms of spin operators $K _{i}$ acting on the $(2k+1)$-dimensional Hilbert space of a particle with spin $k$. In a definite unitary irreducible representation $k$, we built creation and annihilation operators which contract to the usual HW$(1)$ bosonic operators in the large $k$-representation limit. Outstandingly, the corresponding oscillator-like Hamiltonian (which is obtained by replacing the standard bosonic creation and annihilation operators by spin operators) exhibits a maximum number of excitations $2k$ (which is a consequence of the corresponding algebra) and hence the energy spectrum is bounded from above \cite{Chumakov}. The approach adopted here thus leads to a nonlinear extension of the Dicke model that takes into account both an intensity dependent coupling between the atoms and the radiation field, and an additional nonlinear Kerr-like or P\"{o}sch-Teller-like term (depending on the degree of nonlinearity). From now on we will refer to this model as the SU$(2) \, \otimes \,$SU$(2)$ Dicke model, the $k$-Dicke model, or in short the $k$DM. It is worth mentioning that the $k$DM consists of an extension of the dynamical algebra of the system, instead of mutilating directly the bosonic algebra as done within the framework of quantum algebras \cite{Dobrev}. On the other hand, we also emphasize the closeness between our $k$-Dicke model and the two-fluid Lipkin model developed in nuclear physics for the proton-neutron interacting boson model IBM-2 \, \cite{IBM1, IBM2, IBM3}. Quantum-optical examples of two-fluid systems are the Jaynes-Cumming and the Dicke models, for which the two fluids (photons and atoms) play not symmetric roles since they obey different algebras. For a simple two-fluid Lipkin model with SU$(2) \otimes$SU$(2)$ dynamical algebra see Ref.  \onlinecite{PRC}. Upon contraction from SU$(2)$ to HW$(1)$, this particular model reduces to Dicke and Jaynes-Cummings models. The $k$-Dicke model also resembles the Lipkin-Meshkov-Glick model with anisotropic ferromagnetic coupling \cite{SBB}. This model describes an ensemble of all-to-all-coupled two-level systems, whose dynamics has been experimentally explored by using Bose-Einstein condensates. So the results exposed in this work might provide some insight on the behavior of the aforementioned system.

We demonstrate that the most salient feature of the DM, a quantum phase transition from normal to superradiant state, also takes place in the $k$DM. To this end, we use three different approaches: i) mean-field analysis, ii) symmetry-adapted states, and iii) exact computational diagonalization methods. The essential signature of the phase transition threshold value of decay rate is identified in a closed form within the mean field analysis, and then it is corroborated numerically by methods ii) and iii). The threshold for the $k$DM is further showcased to reduce to the well-known threshold in the conventional DM in the large $k$-representation limit as a nature validation. It is worth mentioning that a variety of quantum platforms, such as cavity quantum electrodynamics (QED), trapped ions and circuit QED, provides a natural implementation of the Jaynes-Cummings and Dicke models. Analog-digital quantum simulation of the Dicke Hamiltonian via circuit QED setups provides an interesting arena in which theoretical predictions, in particular those which are impossible to realize in typical cavity QED setups, can be tested. Indeed, circuit QED allow the engineering of a wide range of Kerr-type nonlinearities \cite{Bourassa}, including the one presented in this work. As shown in Ref. \onlinecite{EXP1}, quantum simulations of the Jaynes-Cummings and Tavis-Cummings models contribute to the observation of quantum dynamics not accesible in current experiments. Basically, this approach consists in finding some terms in the simulated system that can be implemented in an analog way, e.g. to employ a harmonic oscillator to simulate a boson field, and focus on the efficient decomposition of the quantum system dynamics in terms of elementary gates. In Ref. \onlinecite{Kirchmair}, for example, the generation of multicomponent Schrodinger-cat-like states was achieved by engineering an artificial Kerr-like medium through circuit QED. Recently, the regular and chaotic behavior of the classical Dicke model have been simulated by using two synthetic non-linearly coupled electric LC oscillators, implemented by means of analog electric components \cite{EXP2}. This technique consists in mapping the original Hamiltonian to an electrical version of the harmonic oscillator, i.e. LC circuits. Interestingly, these techniques (circuit QED and  LC circuits) represent interesting platforms in which the $k$DM can be tested.

The paper is organized as follows. After a brief review of the DM in Sec. \ref{Dicke} we present our algebraic model in Sec. \ref{AlgebraicModel}. There, we discuss the possible applications of our model. The $k$DM is introduced in Sec. \ref{kDickeModel}. In Sec. \ref{QPT}, after discussing the basics of quantum phase transitions, we describe the methodology for calculation we use to investigate the QPT in the $k$DM. Using the suitable trial state, in Sec. \ref{MeanFieldSection} we perform a mean field analysis of the phase transition. Next, in Sec. \ref{ProjectedStatesSection}, we construct parity preserving coherent states (which is possible due to the parity symmetry of the extended Hamiltonian) to analyze also the phase transition. We find that they represent a good approximation to the exact quantum solution of the ground state of the model, which is tackled in Sec. \ref{QuantumSol}. In the concluding section \ref{ConclusionSection} we briefly discuss the possible connections of the $k$DM and the $qp$-deformed algebras.

\section{\label{ModelSection} Model Hamiltonian}

\subsection{\label{Dicke} The Dicke model}

The Dicke Hamiltonian describes $N$ two-level identical atoms, with energy separation $\omega$, interacting collectively with a one-mode radiation field of frequency $\Omega$. In the dipolar approximation, the Hamiltonian (with $\hbar = 1$) reads \cite{Dicke}
\begin{align}
H _{\mbox{\scriptsize D}} = \omega J _{z} + \frac{\Omega}{2} ( a ^{\dagger} a + a a ^{\dagger} ) - \frac{2 \gamma}{\sqrt{N}} J _{x} (a + a ^{\dagger}) , \label{DickeUsual}
\end{align}
where $\gamma$ is the atom-field interaction strength, the collective pseudo-spin operators $J _{i} = \sum _{n = 1} ^{N} S _{i} ^{(n)}$, with $S _{i} = \sigma _{i} / 2$ (being $\sigma _{i}$ the Pauli matrices), satisfy the SU$(2)$ commutation relations
\begin{align}
\left[ J _{+} , J _{-} \right] = 2 J _{z} , \quad \left[ J _{z} , J _{\pm} \right] = \pm J _{\pm} , \label{SU(2)}
\end{align}
and act on a $(2j+1)$-dimensional Hilbert space generated by the Dicke states $\left\lbrace \left| j , m _{j} \right\rangle \right\rbrace$, which are common eigenstates of the  commuting observables $J ^{2}$ and $J _{z}$, with eigenvalues $j(j+1)$ and $m _{j}$, respectively. For the electromagnetic field the creation and annihilation operators $a ^{\dagger}$ and $a$, appearing in the Hamiltonian (\ref{DickeUsual}), satisfy the commutation relations of the Lie algebra generators of the Heisenberg-Weyl group HW$(1)$, i.e.
\begin{align}
\left[ a , a ^{\dagger} \right] = 1 , \quad \left[ a , n \right] = a , \quad \left[ a ^{\dagger} , n \right] = - a ^{\dagger} , \label{HW(1)}
\end{align}
where  $n = a ^{\dagger} a$ is the photon number operator, which acts on an infinite-dimensional Hilbert space generated by the states $\left\lbrace \left| n \right\rangle  \right\rbrace$, with $n$ denoting its corresponding eigenvalue.

\subsection{\label{AlgebraicModel} Algebraic model}

It is well known that the HW$(1)$ algebra can be obtained by contraction of the SU$(2)$ algebra. Here, we consider a new set of pseudo-spin operators $\left\lbrace K _{i} \right\rbrace$ satisfying the SU$(2)$ commutation relations
\begin{align}
\left[ K _{+} , K _{-} \right] = 2 K _{z} , \quad \left[ K _{z} , K _{\pm} \right] = \pm K _{\pm} , \label{KSU(2)}
\end{align}
and, in a definite unitary irreducible representation $k$, we build the operators
\begin{align}
b _{k} = \frac{K _{-}}{\sqrt{2k}} , \quad b ^{\dagger} _{k} = \frac{K _{+}}{\sqrt{2k}} , \quad n _{k} = K _{z} + k , \label{b-operators}
\end{align}
which act on a $(2k+1)$-dimensional Hilbert space of a particle with spin \cite{Chumakov} $k$. In this representation, the SU$(2)$ algebra (\ref{KSU(2)}) becomes
\begin{align}
[ b _{k}, b _{k} ^{\dagger} ] = 1 - \frac{n _{k}}{k} , \quad [ b _{k} , n _{k} ] = b _{k} , \quad [ b ^{\dagger} _{k} , n _{k} ] = - b ^{\dagger} _{k} , \label{b-algebra}
\end{align}
which allows the interpretation of $b _{k}$ and $b ^{\dagger} _{k}$ as the annihilation and creation operators for the quanta labeled by the number operator $n _{k}$. Note that the commutator between the deformed ladder operators ceases to be a $c$-number.

Now we introduce the $k$-oscillator model, which is a quantum system characterized by an oscillator-like Hamiltonian of the form
\begin{align}
H _{k} = \frac{\Omega}{2} (b _{k} b ^{\dagger} _{k} + b ^{\dagger} _{k} b _{k}) ,
\end{align}
where $\Omega$ is a constant. Note that this corresponds to the second term in the usual Dicke Hamiltonian (\ref{DickeUsual}) with the bosonic operators $a$ and $a ^{\dagger}$ replaced by the spin operators $b$ and $b ^{\dagger}$. With the help of the algebra (\ref{b-algebra}), the aforesaid Hamiltonian can equivalently be cast in terms of the number operator as
\begin{align}
H _{k} = \Omega \left( n _{k} + \frac{1}{2} - \frac{n _{k} ^{2} }{2k} \right) . \label{HO-Oscillator}
\end{align}
Note that, in contrast to the harmonic oscillator case, the $k$-oscillator Hamiltonian is no longer a linear function of the number operator. In the limit $k \to \infty$, the $k$-algebra (\ref{b-algebra}) contracts to the HW$(1)$ algebra (\ref{HW(1)}), and the nonlinear term in the Hamiltonian (\ref{HO-Oscillator}) disappears. As we shall discuss in the following, the $k$-oscillator of the type described by the Hamiltonian (\ref{HO-Oscillator}), together with the chosen irreducible representation (\ref{b-operators}), can be used to model two interesting physical systems in different coupling regimes: a Kerr-like medium and a qubit-nonlinear-oscillator system.

It is a well-known fact that a medium exhibits the Kerr effect if its refractive index varies with the intensity of the field. This is the simplest phenomenon of nonlinear optics. Such a system has been recently considered in the framework of the Moyal phase-space representation \cite{Osborn}. On the other hand, Man'ko and coworkers suggested a description within the framework of deformed algebras \cite{Manko}. It is worth pointing out that a close resemblance between the $k$-oscillator Hamiltonian (\ref{HO-Oscillator}) and the usual Kerr Hamiltonian $H _{\mbox{\scriptsize Kerr}} = \Omega (n + 1/2) + \chi n ^{2}$ come up. Here, $\chi$ is an anharmonicity parameter related to the optical properties of the Kerr medium. Usually the Kerr nonlinearity is introduced in a slightly different way: as $a ^{\dagger \, 2} a ^{2} = n ^{2} - n$. This definition introduces a small additional detuning, which nevertheless can be canceled with a redefinition of the parameters. Note that our description of the Kerr medium arises only by relaxing the dynamical algebra of the bosonic field from HW$(1)$ to SU$(2)$, and not by direct deformation of the algebraic structure as done within the quantum algebras. It is worth mentioning that the Kerr-type nonlinearity pushed in this work could be engineered through 3D circuit QED architecture.

To discuss the analogy between the $k$-oscillator and the Kerr medium, it is crucial to observe that the spin excitation number operator $n _{k}$ has nonnegative integer spectrum. The main difference between the photon number operator $n = a ^{\dagger}a$ and the spin excitation number operator $n _{k} = k + K _{z}$ is, however, that the latter is bounded from above: $n _{k} \leq 2k$ (as a consequence of the SU$(2)$ algebra). This result implies that the $k$-oscillator possesses a maximum energy given by $E _{k \, \mbox{\scriptsize max}} = (\Omega / 2) (k+1)$. One can also show that the Heisenberg equation implies that the time evolution of the spin operator $\vec{K} (t)$ is a rotation around the $z$-axis, with the precession frequency depending on the excitation number operator \cite{Chumakov, Martin}. This result is in complete agreement with that obtained in the case of the Kerr medium \cite{Tanas, Milburn}. Notice that the harmonic-oscillator properties are retrieved in the limit $k \to \infty$: $H _{k}$ becomes the usual bosonic oscillator, the spin number of excitations is unbounded and the maximal energy goes to infinity. In this sense, our model certainly describes a weak Kerr medium ($\chi \ll 1$) in the large $k$-representation, determined by $\chi = \Omega /2k$.

The recent development of new technologies have paved the way for implementing hybrid systems incorporating superconducting qubits coupled to oscillator configurations (such as waveguide resonators or SQUID's), admitting the possibility of controlling the parameter settings, on demand, and exploring new quantum phenomena \cite{Xiang, Zhang}. As outstanding solid-state realizations of manageable artificial qubit-oscillator systems we can mention Cooper-pair boxes coupled to superconducting transmission-line resonators \cite{Makhlim} and Josephson flux qubits read out by SQUID's \cite{Chiorescu, Johanson}. In the light of this studies, it makes sense to consider the nonlinear regime of qubit-oscillator configurations. In this regard, it has been recently considered the P\"{o}sch-Teller (PT) oscillator, within the $f$-deformed algebras formalism, to simulate such nonlinear features. Noticeably, the $k$-oscillator we are pushing forward serves also as a full model for the hyperbolic PT oscillator. The relevant Hamiltonian for the hyperbolic P\"{o}sch-Teller oscillator is $H _{\mbox{\scriptsize PT}} = \Omega \left( n + 1/2 - \frac{n ^{2}}{2 \lambda} \right)$, where $\lambda$ is an integer related with the number of bounded states of the system, such that the last bound state corresponds to $n _{\mbox{\scriptsize PT}} = \lambda -1$. This makes our $k$-oscillator model with a maximum number of excitations make sense in qubit-oscillator hybrids.

\subsection{\label{kDickeModel} The $k$-Dicke model}

Now let us generalize the Dicke model writing the Hamiltonian in the following form:
\begin{align}
H _{k\mbox{\scriptsize D}} = \omega J _{z} + \frac{\Omega}{2 k} \left( K ^{2} - K _{z} ^{2} \right) - \frac{4 \gamma}{\sqrt{2 k N}} J _{x} K _{x} ,  \label{kDicke}
\end{align}
which is obtained by replacing the bosonic ladder operators $a$ and $a ^{\dagger}$ by the spin operators $b$ and $b ^{\dagger}$ in the usual Dicke Hamiltonian (\ref{DickeUsual}). We will henceforth refer to $H _{k\mbox{\scriptsize D}}$ as the $k$-Dicke Hamiltonian. Note that $H _{k\mbox{\scriptsize D}}$ manifests now that the dynamical algebra of the extended model is SU$(2) \, \otimes \,$SU$(2)$. It is worth mentioning that the $k$-Dicke Hamiltonian (9) exhibits some similarities with that of the two-fluid Lipkin model presented in Ref. \onlinecite{PRC}. This Hamiltonian resembles the proton-neutron interacting boson model Hamiltonian of interest in nuclear physics \cite{IBM1, IBM2, IBM3}. Interestingly, the consistent $Q$-like double Lipkin Hamiltonian of Ref. \onlinecite{PRC} would exhibit a mixture of dynamical symmetries for different choices of the parameter $x$, thus describing transitions between spherical and deformed phases.

As in the Dicke Hamiltonian (\ref{DickeUsual}), the first term in Eq. (\ref{kDicke}) is the energy operator for the atoms. The second term corresponds to the energy operator for the field, which in this case corresponds to the model Hamiltonian (\ref{HO-Oscillator}) expressed in terms of the $\left\lbrace K _{i} \right\rbrace$ pseudo-spin operators. In the dipole approximation, the coupling between the ensemble of atoms and the finite radiation field is given by the last term above. As we can see in (\ref{kDicke}), the essential feature of our model is twofold. On the one hand, it is associated with an intensity dependent coupling between the atoms and the radiation field (i.e. the coupling is no longer linear in the field variables). On the other hand, the Hamiltonian for the radiation field, $H _{k}$, may be in turn interpreted as a subsystem that governs the behavior of the field surrounded by a nonlinear medium inside the cavity. As discussed in the previous section, it could be a weak Kerr-like medium or a P\"{o}sch-Teller type nonlinear-oscillator. Finally, one can further see that the Hamiltonian $H _{k\mbox{\scriptsize D}}$ reduces to $H _{\mbox{\scriptsize D}} + \Omega /2$ under the group contraction (i.e. in the limit $k \to \infty$), as it should be. Also, the Hamiltonian (\ref{kDicke}) commutes with the operators $J ^{2}$ and $K ^{2}$. Consequently, it connects only states with the same total spins $j$ and $k$, i.e. that belong to the same Dicke manifold. Let us keep $k$ finite and investigate the physical properties of the extended Dicke model (\ref{kDicke}). 

It is worth underlying that in a foregoing investigation we introduced the $k$-extended Jaynes-Cumming model (in the dipole and rotating wave approximations), which remains exactly solvable with the new SU$(2) \, \otimes \,$SU$(2)$ dynamical algebra \cite{Martin}. There we showed that the temporal evolution of both the atomic and field quantum properties (e.g. collapses and revivals, photon anti-bunching and squeezing) exhibit significant different behavior from that of the usual JC model for small values of $k$.

\section{\label{QPT} Quantum Phase Transitions}

\subsection{\label{Def} Definition}

Phase transitions, both quantum and classical, can be informally thought as sudden, drastic changes in the properties of a system due to the variation of some parameter relevant to it. In the particular case of quantum phase transitions (QPTs), this change is produced in the ground state of the system at zero temperature. Of course, it is impossible to cool any system down to $T = 0K$, however, as it has been experimentally shown in some works \cite{key-5,key-6}, at finite temperatures close to the absolute zero, in the regime where $\hbar \omega \gg k_{B} T$, the phenomena described by the QPTs can be observed.

A ``quantum phase'' is defined formally as an open region $\mathcal{D}\subseteq\mathbb{R}^{\ell}$	where the ground state's energy $\mathcal{E}_{0}$, as a function of $\ell$ parameters involved in the modeling Hamiltonian, is analytic. A QPT is then identified by the boundary $\partial\mathcal{D}$ of the region at which $\frac{\partial^{n}\mathcal{E}_{0}}{\partial x^{n}}$ is discontinuous for some $n$ (known as the order of the transition). This is precisely the definition of QPT we shall use in this paper. 

It is worth mentioning that this definition is not universal. It is somewhat common to refer to this concept as a quantum crossover instead of a quantum phase transition, reserving the latter for a quantum crossover that remains in the thermodynamic limit.

\subsection{\label{Met} Methodology for calculation}

In order to know if the system modeled by Hamiltonian (\ref{kDicke}) undergoes a quantum phase transition, we use the energy surface minimization method, which consists of minimizing the surface that is obtained by taking the expectation value of the Hamiltonian with respect to some trial variational state. This method allows us to approximate the ground state of the system and has been extensively used to study the phase transition in the Dicke model \cite{Castanos1, Castanos2, Castanos3, Castanos4, Castanos5, Nahmad1, Quezada1} as well as in more general models using three-level systems \cite{Cordero1, Cordero2, Cordero3, Cordero4, Quezada2}. Of course, the success of the method strongly depends on the choice of the trial state used to model the ground state of the system.

In this work, we first perform a mean field analysis by using the tensor product of the coherent states in each subspace as a trial state. Next, we employ the symmetry-adapted states, which are parity-preserving trial states obtained by projecting the tensor product of coherent states. Finally, we obtain the exact quantum solution by numerically diagonalizing the Hamiltonian (\ref{kDicke}). The motivation behind choosing these specific trial variational states is that, in the Dicke model, all these approaches (mean-field, symmetry-adapted and numerical) converge in the thermodynamic limit, leading to a well-defined, second-order quantum phase transition \cite{Nahmad1, Quezada1}. On the other hand, across a quantum phase transition, the ground state of a system suffers a sudden, drastic change. Thus it is natural to use the fidelity between neighboring states to detect a QPT. 

For a pure quantum state $\left| \psi \right>$, the fidelity between neighboring states is defined as
\begin{align}
\mathcal{F} = \vert \left< \psi (\gamma)  \vert \psi (\gamma + \delta \gamma ) \right> \vert ^{2}, \label{fidelity}
\end{align}
and it measures the overlap between the state at $\gamma$ and the same state at \cite{Zanardi} $\gamma + \delta \gamma$. In this work we use the fidelity between neighboring states to detect the QPT in the exact quantum solution (the numerical approach) of the $k$-Dicke model, as this method has already been proven useful for the same task in the Dicke model \cite{Nahmad1, Quezada1, Bastarrachea}.

\subsection{\label{QPTD} QPT in the Dicke Model}

The Dicke Hamiltonian (\ref{DickeUsual}) undergoes a second-order quantum phase transition at a critical value of the atom-field coupling strength $\gamma _{c} ^{D} = \sqrt{\omega N \Omega / 8j}$. The system transits from a normal phase ($\gamma < \gamma _{c} ^{D}$), where the spontaneous radiation rate of the system is proportional to the number of atoms $N$, to a super-radiant phase \cite{Hepp-Lieb, Hepp-Lieb2, Wang-Hioe} ($\gamma > \gamma _{c} ^{D}$), where the spontaneous radiation rate of the system is proportional to the squared number of atoms \cite{Dicke} $N^{2}$. It is worth mentioning that this critical value can be obtained via the surface minimization method by using the tensor product of the coherent states of both matter and field \cite{Nahmad1, Quezada1}, furthermore, in the case $j = \frac{N}{2}$, the transition remains in the thermodynamic limit.

\section{\label{MeanFieldSection} Mean field analysis}

%%%%%%%%%%%%%%%%%%%%%%%%%%%%%%%%%%%%%%%%%%%%%%%%%%%%%%%%%%%%%%%%%%%%%%%%%%
\begin{figure*}[t]
	a\includegraphics[width = 2.7in]{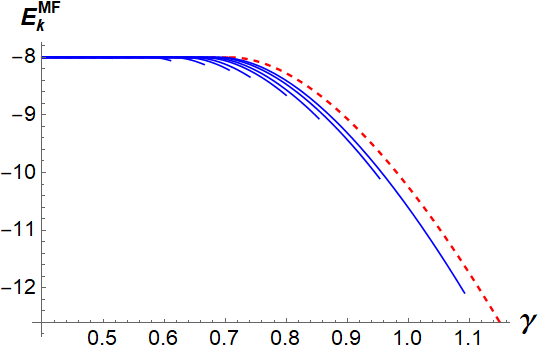} \qquad \qquad \qquad
	b\includegraphics[width = 2.7in]{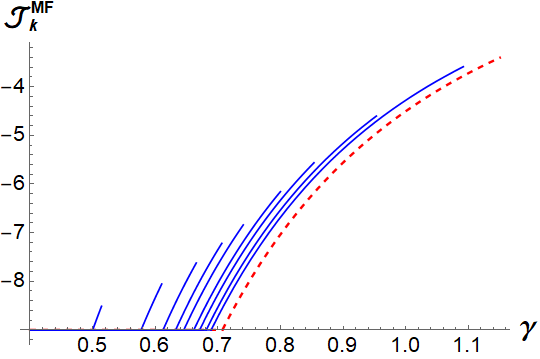}
	c\includegraphics[width = 2.7in]{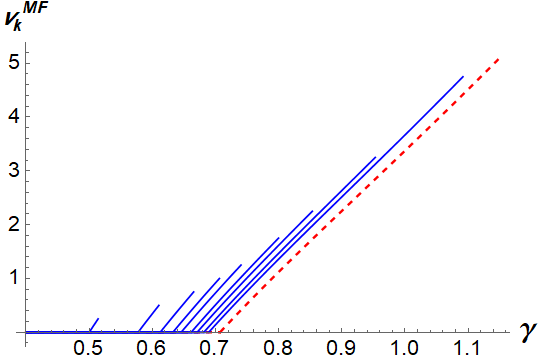}
	'\caption{\small Plot of the ground-state energy $E _{k} ^{\mbox{\tiny MF}}$ (a), the expectation value of the atomic relative population operator $\mathcal{J} _{k} ^{\mbox{\tiny MF}}$ (b) and the expectation value of the analogue of the photon number operator $\nu _{k} ^{\mbox{\tiny MF}}$ (c), as function of the field-matter coupling $\gamma$ for $k\in \left\lbrace 1,\frac{3}{2},2,\frac{5}{2},3,4,5,7,10 \right\rbrace$ (blue-continuous lines). As a benchmark, the red-dashed lines show these same quantities in the limiting case $k \to \infty$, which corresponds to the HW$(1) \, \otimes \,$SU$(2)$ Dicke model. All plots were obtained using $\omega = 1$, $\Omega = 2$, $N = 18$ and $j = 9$.} \label{PlotsCoherent}
\end{figure*}
%%%%%%%%%%%%%%%%%%%%%%%%%%%%%%%%%%%%%%%%%%%%%%%%%%%%%%%%%%%%%%%%%%%%%%%%%%

In order to obtain an energy surface, we first use as a trial state the direct product of coherent states in each subspace. For the matter sector, we use the standard SU$(2)$ spin states \cite{Arecchi}, i.e.
\begin{align}
\left| \xi \right> = \frac{1}{\left( 1 + \vert \xi \vert ^{2} \right) ^{j}} \sum _{m _{j} = 0} ^{2j} \binom{2j}{m _{j}} ^{1/2} \xi ^{m _{j}} \left| j, m _{j} - j \right> ,
\end{align}
where $\xi = \tan (\theta / 2) e ^{i \phi}$, and $\left| j, m _{j} \right>$ is a Dicke state. The angles $\theta \in [0, \pi )$ and $\phi \in [0, 2 \pi )$ determine a point on the Bloch sphere. In the problem at hand, the radiation field is modeled by the $k$-oscillator, and thus we use the SU$(2)$ spin coherent state,
\begin{align}
\left| \eta \right> = \frac{1}{\left( 1 + \vert \eta \vert ^{2} \right) ^{k}} \sum _{m _{k} = 0} ^{2k} \binom{2k}{m _{k}} ^{1/2} \eta ^{m _{k}} \left| k, m _{k} - k \right> ,
\end{align}
for the field sector, which contracts to the Heisenberg-Weyl coherent state $\left| \alpha \right\rangle$ in the large $k$-representation limit \cite{Martin, Eberly}. Here, $\eta = \tan (\psi / 2) e ^{i \varphi}$, $\psi \in [0, \pi )$ and $\varphi \in [0, 2 \pi )$. The trial state $\left| \Psi \right> = \left| \xi \right> \otimes \left| \eta \right>$ contains $N$ particles and up to $2k$ photons distributed in all possible ways between the two levels.

The expectation value of the $k$-Dicke Hamiltonian in this trial state is defined as the energy surface, i.e. $E _{k} ^{\mbox{\tiny MF}} ( \vec{\lambda} ) \! = \! \left< \Psi \right| H _{k\mbox{\scriptsize D}} \left| \Psi \right> \! = \! \left< H _{k\mbox{\scriptsize D}} \right> ^{\mbox{\tiny MF}}$, where $ \vec{\lambda} = (\theta , \phi , \psi , \varphi)$. The superscript MF indicates that the expectation value is calculated in the state $\left| \Psi \right>$, within the mean field analysis. Straightforward calculations show that the energy surface can be written in the simple form
\begin{align}
E _{k} ^{\mbox{\tiny MF}} ( \vec{\lambda} ) &= \omega \mathcal{J} _{k} ^{\mbox{\tiny MF}} (\vec{\lambda}) +  \Omega \left( \nu _{k} ^{\mbox{\tiny MF}} (\vec{\lambda}) + \frac{1}{2} \right) + \gamma \, \mathcal{I} _{k}  ^{\mbox{\tiny MF}} (\vec{\lambda}) , \label{EnergySurface}
\end{align}
where
\begin{align}
\mathcal{J} _{k} ^{\mbox{\tiny MF}} (\vec{\lambda}) \equiv \left\langle J _{z} \right\rangle ^{\mbox{\tiny MF}} = - j \cos \theta  \label{Jz} 
\end{align}
is the expectation value of the atomic relative population operator, 
\begin{align}
\mathcal{I} _{k} ^{\mbox{\tiny MF}} (\vec{\lambda}) & \equiv - \frac{4}{\sqrt{2kN}} \left< J _{x} K _{x} \right> ^{\mbox{\tiny MF}} \notag \\ &=  -2j \sqrt{\frac{2k}{N}} \sin \theta \sin \psi \cos \phi \cos \varphi  \label{Inter}
\end{align}
is the expectation value of the interaction term in the Hamiltonian (\ref{kDicke}), and
\begin{align}
\nu _{k} ^{\mbox{\tiny MF}} (\vec{\lambda}) \equiv \left\langle \frac{H _{k}}{\Omega} - \frac{1}{2} \right\rangle ^{\mbox{\tiny MF}} \!\! = \frac{1}{2} (k - 1/2) \sin ^{2} \psi . \label{nk} 
\end{align} 
Notice that $\nu _{k} ^{\mbox{\tiny MF}}(\vec{\lambda})$ is defined dividing the total energy of the radiation field in equal chunks of energy $\Omega$ and then subtracting the zero-point energy, which is clearly different from $\left\langle n_k \right\rangle ^{\mbox{\tiny MF}}$ with $n_k$ as in eq. \eqref{b-operators}. Indeed one can further verify that $\nu _{k} ^{\mbox{\tiny MF}} = \left\langle n_k \right\rangle ^{\mbox{\tiny MF}} - \frac{1}{2k} \left\langle n_k ^{2} \right\rangle ^{\mbox{\tiny MF}}$, as it should be. The motivation behind the definition of $\nu _{k} ^{\mbox{\tiny MF}}$ is only to maintain a close analogy with the expectation value of the photon number operator in the usual Dicke model, although they are physically different. In Sec. \ref{ProjectedStatesSection}, where the total excitation number operator is required to construct symmetry-adapted states, we use certainly the number operator $n_k$ which correctly closes the SU(2) algebra.

The critical points $\vec{\lambda} _{c} ^{\mbox{\tiny MF}}$ of the energy surface (\ref{EnergySurface}) are obtained by equating its first derivative  to zero, i.e. 
\begin{align}
\frac{\partial E _{k} ^{\mbox{\tiny MF}} (\vec{\lambda})}{\partial \vec{\lambda}} \Big| _{\vec{\lambda} = \vec{\lambda} _{c} ^{\mbox{\tiny MF}}} = \vec{0} .
\end{align}
Those which correspond to minima in the energy surface are
\begin{align}
\vec{\lambda} _{1} ^{\mbox{\tiny MF}} = (0 , \phi _{0} , 0 , \varphi _{0} ) , \quad \mbox{for} \quad \vert \gamma \vert < \gamma _{c} ,
\end{align}
being $\phi _{0}$ and $\varphi _{0} $ arbitrary azimuthal angles; and
\begin{align}
\vec{\lambda} _{2} ^{\mbox{\tiny MF}} = ( \theta _{c} , \phi _{c} , \psi _{c} , \varphi _{c} ) , \quad \mbox{for} \quad  \gamma _{c} < \vert \gamma \vert < \gamma _{m},
\end{align}
where
\begin{align}
\theta _{c} &= \arccos (\gamma _{c} / \gamma ) ^{2} , \label{CriticalTheta} \\ \psi _{c} &= \arcsin \left[ \frac{\mu \, \omega \, \gamma}{2 \gamma _{c} ^{2}} \sqrt{\frac{N}{2k}} \sqrt{1 - (\gamma _{c} / \gamma ) ^{4}} \right] , \label{CriticalPsi}
\end{align}
with $\mu \equiv \cos \varphi _{c} \cos  \phi _{c} = \pm 1$. There are four solutions to the later condition in the domain $[0, 2 \pi ) \times [0 , 2 \pi )$. They are $(0,0)$, $(\pi , \pi)$, $(0, \pi)$ and $(\pi , 0)$, with the notation $(\phi _{c} , \varphi _{c})$. Since the sign of $\mu$ can be absorbed into the definition of $\gamma$, both cases will produce exactly the same physics, as we shall see below. In eqs. (\ref{CriticalTheta}) and (\ref{CriticalPsi}), we have defined the critical value of the field-matter coupling,
\begin{align}
\gamma _{c} = \sqrt{\frac{\omega N \Omega (k-1/2)}{8jk}} ,
\end{align}
at which a phase transition occurs, and from the fact that max$(\sin \psi _{c}) = 1$ in the domain $[0, \pi)$ we obtain a cut-off value for the coupling
\begin{align}
\gamma _{m} = \gamma _{c} \; \sqrt{\frac{\Omega (k-1/2)}{2 \omega j} + \sqrt{1 + \left[ \frac{\Omega (k-1/2)}{2 \omega j} \right] ^{2}}} , \label{MaximumCoupling}
\end{align}
which is a direct consequence of the maximum number of excitations of the $k$-oscillator. The condition $ \vert \gamma \vert < \gamma _{c}$ defines the normal phase (where the ground state has zero photons and no excited atoms) and the condition $\gamma _{c} < \vert \gamma \vert  < \gamma _{m}$ describes the super-radiant phase. As we shall see below, a phase transition occurs at $\gamma = \gamma _{c}$. Note that, as expected, the large $k$-representation limit ($k \to \infty$) leads to $\gamma _{c} \to \gamma _{c} ^{D} = \sqrt{\omega N \Omega / 8j}$ (the critical value of the usual Dicke model) and $\gamma _{m} \to \infty$ (unbounded number of photons).

Substituting the critical points $\vec{\lambda} _{1} ^{\mbox{\tiny MF}}$ and $\vec{\lambda} _{2} ^{\mbox{\tiny MF}}$ into eq. \eqref{EnergySurface} we obtain the energy of the coherent ground state as a function of the Hamiltonian parameters,
\begin{align}
E _{k} ^{\mbox{\tiny MF}} ( \gamma ) \! = \! \left\lbrace \begin{array}{l} \!\! \frac{\Omega}{2} \! - \! j \omega , \\[7pt] \!\! \frac{\Omega}{2} \! - \! \frac{j \omega}{2} \! \left[ \left( \gamma / \gamma _{c} \right) ^{2} \!\! + \! \left( \gamma _{c} / \gamma \right) ^{2} \right] \! , \end{array} \right.  \!\!\!\! \begin{array}{l} \mbox{for} \; \vert \gamma \vert \! < \! \gamma _{c}  \\[7pt] \mbox{for} \;  \gamma _{c} \! < \! \vert \gamma \vert  \! < \! \gamma _{m} \end{array} \!\! . \label{Ecoh}
\end{align}
In a similar fashion, we can also obtain that the expectation values of the atomic relative population operator and the photon number operator can be written as 
\begin{align}
\mathcal{J} _{k} ^{\mbox{\tiny MF}} (\gamma) \! = \! \left\lbrace \begin{array}{l} - j , \\[7pt]  - j \! \left( \gamma _{c} / \gamma \right) ^{2} \!, \end{array} \right.  \begin{array}{l} \mbox{for} \; \vert \gamma \vert \! < \! \gamma _{c}  \\[7pt] \mbox{for} \;  \gamma _{c} \! < \! \vert \gamma \vert  \! < \! \gamma _{m} \end{array} \!\! , \label{Jzcoh}
\end{align}
and
\begin{align}
\nu _{k} ^{\mbox{\tiny MF}} (\gamma)\! = \! \left\lbrace \begin{array}{l} 0  \; ,  \\[7pt]  \!\! \frac{j \omega}{2 \Omega} \left[ (\gamma / \gamma _{c} ) ^{2} \! - \! (\gamma _{c} / \gamma ) ^{2}  \right] , \end{array}  \right. \!\! \begin{array}{l} \mbox{for} \,\, \vert \gamma \vert < \gamma _{c} \\ [7pt] \mbox{for} \,\, \gamma _{c} < \vert \gamma \vert < \gamma _{m} \end{array} . \label{PhotonNumber}
\end{align}
respectively. In Fig. \ref{PlotsCoherent} we show the ground-state energy $E _{k} ^{\mbox{\tiny MF}}$ (at left), the expectation values of the atomic relative population operator $\mathcal{J} _{k} ^{\mbox{\tiny MF}}$ (at center) and the photon number operator $\nu _{k} ^{\mbox{\tiny MF}}$ (at right), as a function of the field-matter coupling $\gamma$ for $N = 2j = 18$ and $k\in \left\lbrace 1,\frac{3}{2},2,\frac{5}{2},3,4,5,7,10 \right\rbrace$. As a benchmark, the red-dashed lines show these same quantities for the archetypal Dicke model. We observe that the energy, atomic population and photon number in the $k$-Dicke model, approaches asymptotically to the usual results for the HW$(1) \, \otimes \,$SU$(2)$ Dicke model as increasing $k$, as expected. For small values of $k$, we observe significant departures from the usual results mainly because both the critical value $\gamma _{c}$ and the cutoff value $\gamma _{m}$ of the field-matter coupling decreases. As we will show below, the same profiles are obtained for the energy, the atomic relative population operator and the photon number operator, when computed with symmetry-adapted variational states as well as for exact quantum solutions.

\section{\label{ProjectedStatesSection} Symmetry-adapted states (SAS) analysis}

%%%%%%%%%%%%%%%%%%%%%%%%%%%%%%%%%%%%%%%%%%%%%%%%%%%%%%%%%%%%%%%%%%%%%%%%%%
\begin{figure*}[t]
	a\includegraphics[width = 2.7in]{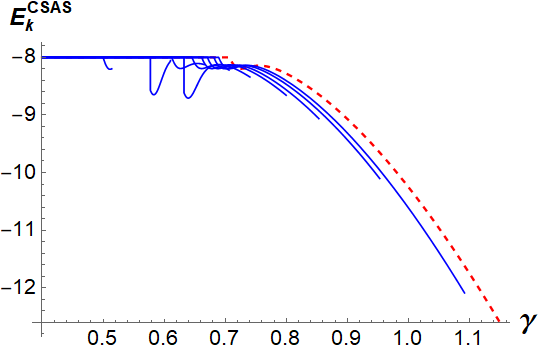} \qquad \qquad \qquad
	b\includegraphics[width = 2.7in]{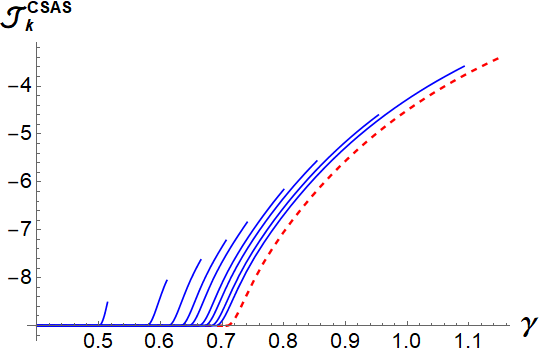}
	c\includegraphics[width = 2.7in]{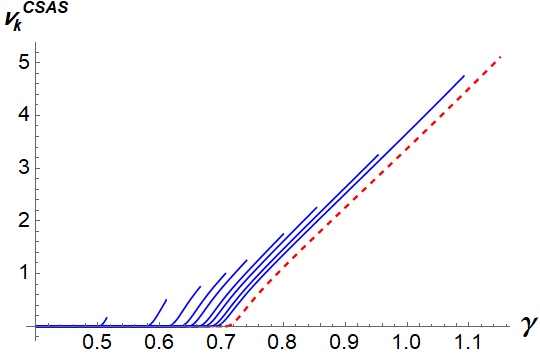}
	'\caption{\small Plot of the ground-state energy $E _{k} ^{\mbox{\tiny CSAS}}$ (a), the expectation value of the atomic relative population operator $\mathcal{J} _{k} ^{\mbox{\tiny CSAS}}$ (b) and the expectation value of the analogue of the photon number operator $\nu _{k} ^{\mbox{\tiny CSAS}}$ (c) as a function of the field-matter coupling $\gamma$, obtained using the MF critical points in the SAS expressions, for $k\in \left\lbrace 1,\frac{3}{2},2,\frac{5}{2},3,4,5,7,10 \right\rbrace$ (blue-continuous lines). As a benchmark, the red-dashed lines show these same quantities in the limiting case $k \to \infty$, which corresponds to the HW$(1) \, \otimes \,$SU$(2)$ Dicke model. All plots were obtained using $\omega = 1$, $\Omega = 2$, $N = 18$ and $j = 9$.} \label{PlotsSAS}
\end{figure*}
%%%%%%%%%%%%%%%%%%%%%%%%%%%%%%%%%%%%%%%%%%%%%%%%%%%%%%%%%%%%%%%%%%%%%%%%%%

%%%%%%%%%%%%%%%%%%%%%%%%%%%%%%%%%%%%%%%%%%%%%%%%%%%%%%%%%%%%%%%%%%%%%%%%%%
\begin{figure*}[t]
	a\includegraphics[width = 2.7in]{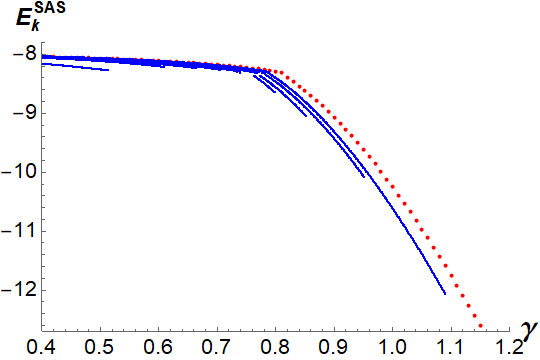} \qquad \qquad \qquad
	b\includegraphics[width = 2.7in]{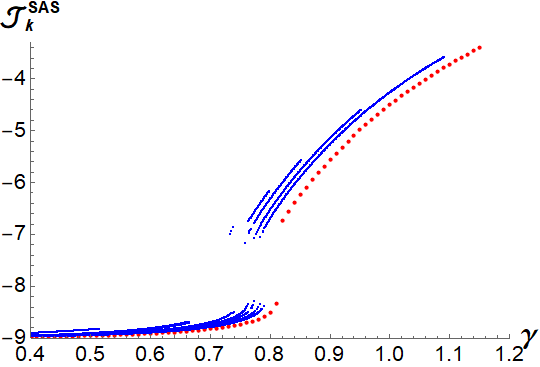}
	c\includegraphics[width = 2.7in]{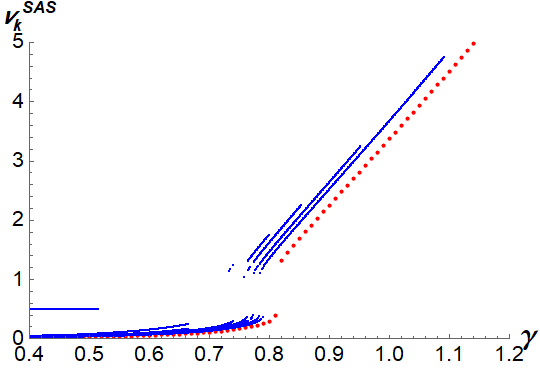}
	'\caption{\small Plot of the ground-state energy $E _{k} ^{\mbox{\tiny SAS}}$ (a), the expectation value of the atomic relative population operator $\mathcal{J} _{k} ^{\mbox{\tiny SAS}}$ (b) and the expectation value of the analogue of the photon number operator $\nu _{k} ^{\mbox{\tiny SAS}}$ (c) as a function of the field-matter coupling $\gamma$, obtained by numerically minimizing the SAS energy surface, for $k\in \left\lbrace 1,\frac{3}{2},2,\frac{5}{2},3,4,5,7,10 \right\rbrace$ (blue-continuous lines). As a benchmark, the red-dashed lines show these same quantities in the limiting case $k \to \infty$, which corresponds to the HW$(1) \, \otimes \,$SU$(2)$ Dicke model. All plots were obtained using $\omega = 1$, $\Omega = 2$, $N = 18$ and $j = 9$.} \label{PlotsSASNum}
\end{figure*}
%%%%%%%%%%%%%%%%%%%%%%%%%%%%%%%%%%%%%%%%%%%%%%%%%%%%%%%%%%%%%%%%%%%%%%%%%%

The $k$-Dicke Hamiltonian (\ref{kDicke}) has a parity symmetry given by
\begin{align}
\left[ e ^{i \pi \Lambda} , H _{k\mbox{\scriptsize D}} \right] = 0 , \label{Symmetry}
\end{align}
where
\begin{align}
\Lambda = \sqrt{J ^{2} + \frac{1}{4}} - \frac{1}{2} + J _{z} + \sqrt{K ^{2} + \frac{1}{4}} - \frac{1}{2} + K _{z}  \label{NumberExcOp}
\end{align}
is the excitation number operator with eigenvalues $\lambda =  j + m _{j} + k + m _{k}$. This allows for the classification of the eigenstates of $H _{k\mbox{\scriptsize D}}$ in terms of the parity of the eigenvalues $\lambda$. We observe that the set of SU$(2)$ coherent states $\left\lbrace \left| \xi \right> , \left| \eta \right> \right\rbrace$ strongly mixes states with different parity. Therefore, we can build up symmetry-adapted coherent states that preserve the symmetry of the $k$-Dicke Hamiltonian, by acting with the projectors of the symmetric and anti-symmetric representations of the cyclic group  \cite{Castanos} $C _{2}$
\begin{align}
\mathcal{P} _{\pm } = \frac{1}{2} \left( I \pm e ^{i \pi \Lambda} \right) , \label{Projectors}
\end{align}
upon the state $\left| \Psi \right> = \left| \xi \right> \otimes \left| \eta \right>$. Explicitly we have
\begin{align}
\left| \xi , \eta \right> _{\pm} &= \mathcal{N} _{\pm} \mathcal{P} _{\pm} \left| \xi  \right> \otimes \left| \eta \right> \notag \\ &= \mathcal{N} _{\pm}  \left( \left| \xi  \right> \otimes \left| \eta \right>\pm  \left| - \xi  \right> \otimes \left| - \eta \right> \right) , \label{SAS}
\end{align}
where the normalization factor for the even (+) and odd (-) states are
\begin{align}
\mathcal{N} _{\pm} ( \theta , \psi ) = \left[2  \pm 2 \left( - \cos \theta \right) ^{2j}  \left( - \cos \psi \right) ^{2k} \right] ^{-1/2}  . \label{NormalizationSAS}
\end{align}
Now we perform a variational analysis of the problem by using the symmetry-adapted states (\ref{SAS}) as a trial state. As before, the energy surface is defined as the expectation value of the $k$-Dicke Hamiltonian in the state $\left| \xi , \eta \right> _{\pm}$, i.e. $E _{k , \pm} ^{\mbox{\tiny SAS}} (\vec{\lambda}) = {}_{\pm} \! \left< \xi , \eta \right| H _{k\mbox{\scriptsize D}} \left| \xi , \eta \right> _{\pm}$. The energy surface can be written once more as
\begin{align}
E _{k , \pm} ^{\mbox{\tiny SAS}} (\vec{\lambda}) =& \,  \omega  \mathcal{J} _{k, \pm}  ^{\mbox{\tiny SAS}} (\vec{\lambda}) + \Omega \left( \nu _{k, \pm}  ^{\mbox{\tiny SAS}} (\vec{\lambda}) + 1/2 \right) + \gamma \, \mathcal{I} _{k, \pm}  ^{\mbox{\tiny MF}} (\vec{\lambda} ) , \label{EnergySASfin}
\end{align}
where the expectation values appearing in this expression (computed in the SAS) can be expressed as
\begin{align}
\mathcal{J} _{k, \pm}  ^{\mbox{\tiny SAS}} (\vec{\lambda}) \! &= - j \cos \theta  \,  \frac{1 \pm (\cos \theta) ^{2j-2} (- \cos \psi) ^{2k}}{1 \pm ( - \cos \theta) ^{2j} (- \cos \psi) ^{2k}} , \label{ExpValuesSAS} \\ \nu _{k, \pm}  ^{\mbox{\tiny SAS}} (\vec{\lambda}) \! &= \! \frac{k \!- \!1/2}{2}  \! \left[ 1 \! - \! \frac{\cos ^{2} \! \psi \pm \! ( - \cos \theta) ^{2j} (- \cos \psi) ^{2k-2} }{1 \pm ( - \cos \theta) ^{2j} (- \cos \psi) ^{2k}} \right] \!\! , \notag \\ \mathcal{I} _{k, \pm}  ^{\mbox{\tiny SAS}}  (\vec{\lambda}) \! &= -2j \sqrt{\frac{2k}{N}} \sin \theta \sin \psi \cos \phi \cos \varphi  \notag \\ & \phantom{=} \times  \frac{1 \mp \tan \phi \tan \varphi ( - \cos \theta) ^{2j-1} (- \cos \psi) ^{2k-1} }{1 \pm ( - \cos \theta) ^{2j} (- \cos \psi) ^{2k}} . \notag
\end{align}
Now we have to minimize the energy surface (\ref{EnergySASfin}) by requiring its derivative to be zero at the critical points $\vec{\lambda} _{c} ^{\mbox{\tiny SAS}}$. This is a difficult task in this case mainly due to the intricate angular dependence of the above expectation values. However, as a first approximation, we may substitute the critical points $\vec{\lambda} _{1} ^{\mbox{\tiny MF}}$ and $\vec{\lambda} _{2} ^{\mbox{\tiny MF}}$ obtained for the mean field energy surface (\ref{EnergySurface}), a method we will refer to as the CSAS approach. Indeed, one can further verify that the critical point $\vec{\lambda} _{1} ^{\mbox{\tiny MF}}$ is still a critical point of the SAS energy surface, i.e. $\vec{\lambda} _{1} ^{\mbox{\tiny SAS}} = \vec{\lambda} _{1} ^{\mbox{\tiny MF}}$. A simple calculation shows that the expectation value of the atomic relative population operator is
\begin{align}
\mathcal{J} _{k, \pm}  ^{\mbox{\tiny CSAS}} ( \gamma ) \! &= \left\lbrace \begin{array}{l} \!\!\! - j \; \frac{1 \pm (- 1) ^{2k}}{1 \pm ( - 1) ^{2(j+k)}} , \\[8pt] \!\!\! - j \lambda ^{2}  \,  \frac{1 \pm (- 1) ^{2k} \lambda ^{4(j-1)} \Gamma ^{k}}{1 \pm ( - 1) ^{2(j+k)} \lambda ^{4j} \Gamma ^{k}}  , \end{array} \right. \begin{array}{l} \mbox{for} \; \vert \gamma \vert \! < \! \gamma _{c}  \\[7pt] \mbox{for} \;  \gamma _{c} \! < \! \vert \gamma \vert  \! < \! \gamma _{m} \end{array} \!\! ,
\end{align}
where $\Gamma \equiv 1 + \delta ( \lambda ^{2} - \lambda ^{-2} )$, $\lambda \equiv \gamma _{c} / \gamma $ and $\delta \equiv \frac{j \omega}{\Omega (k-1/2)}$ are dimensionless parameters. In a similar fashion, we find that the expectation value of the photon number operator is $\nu _{k} ^{\mbox{\tiny CSAS}} (\gamma) = 0$ for $\vert \gamma \vert < \gamma _{c}$, and 
\begin{align}
\nu _{k} ^{\mbox{\tiny CSAS}} (\gamma)\! = \frac{k-1/2}{2} \! \left[\! 1 \! - \! \frac{\Gamma \pm (-1) ^{2(j+k)} \lambda ^{4j} \Gamma ^{k-1}}{1 \pm (-1) ^{2(j+k)} \lambda ^{4j} \Gamma ^{k}} \right] \label{PhotonNumberSAS}
\end{align}
for $\gamma _{c} < \vert \gamma \vert  < \gamma _{m}$.

From the above results we infer that the energy in the normal phase $\vert \gamma \vert < \gamma _{c}$ is a constant, namely
\begin{align}
E _{k , \pm} ^{\mbox{\tiny CSAS}} (\gamma) = \frac{\Omega}{2} - j \omega \frac{1 \pm (- 1) ^{2k}}{1 \pm ( - 1) ^{2(j+k)}} . \label{EnergySASnormal}
\end{align}
In the usual Dicke model, the ground state of the system has an even parity for an integer $j$, as considered in Ref. \onlinecite{Quezada1}, and an odd parity for a half-integer $j$. In the $k$-Dicke model, which is described by two interacting fermion fields, the parity of the ground state depends on both spins. For example, as we read from eq. \eqref{EnergySASnormal}, an even parity ground state with $k$ and $j$ both integers produces $- j \omega + \frac{\Omega}{2}$. However, the same result is obtained with an odd parity ground state with integer $j$ and half-integer $k$.

The energy in the super-radiant phase $\gamma _{c} < \vert \gamma \vert <  \gamma _{m}$ is
\begin{align}
E _{k , \pm} ^{\mbox{\tiny CSAS}} (\gamma) &= \frac{\Omega}{2} - j \omega \lambda ^{2}  \,  \frac{1 \pm (- 1) ^{2k} \lambda ^{4(j-1)} \Gamma ^{k}}{1 \pm ( - 1) ^{2(j+k)} \lambda ^{4j} \Gamma ^{k}} \notag \\ &+ \frac{\Omega}{2} \left( k-1/2 \right) \! \left[\! 1 \! - \! \frac{\Gamma \pm (-1) ^{2(j+k)} \lambda ^{4j} \Gamma ^{k-1}}{1 \pm (-1) ^{2(j+k)} \lambda ^{4j} \Gamma ^{k}} \right] \notag \\ &- \frac{j \omega}{\lambda^{2}} \frac{1 - \lambda ^{4}}{1 \pm (-1)^{2(j+k)} \lambda ^{4j} \Gamma ^{k}} . \label{EnergySASsuper}
\end{align}

In figure \ref{PlotsSAS} we plot the ground state energy (at left), the expectation value of the atomic relative population operator (at middle) and the expectation value of the excitation number operator (at right), all for $N = 2j = 18$ (integer $j$) and $k\in \left\lbrace 1,\frac{3}{2},2,\frac{5}{2},3,4,5,7,10 \right\rbrace$, using the odd-parity ground state for $k = \frac{3}{2}, \frac{5}{2}$ and the even-parity ground state for $k = 1, 2, 3, 4, 5, 7, 10$. It can be observed that the behavior of the analyzed quantities is slightly different from the mean-field expressions in equations \eqref{Ecoh}, \eqref{Jzcoh} and \eqref{PhotonNumber}. Mainly, the energy plot shows an oscillatory-like behavior near the transition, with a higher amplitude for the odd-parity case. In general, all quantities appear to have a smoother transition than its analogous mean-field plots in figure \ref{PlotsCoherent}.

Another way to approximate the ground state of the system is to numerically minimize the SAS energy surface in eq. \eqref{EnergySASfin}. This approach has the disadvantage of not letting us have analytical expressions for the energy, the expectation value of the atomic relative population operator and the expectation value of the excitation number operator, nevertheless it allows us to have a better understanding of the behavior of these quantities in the normal region, where the MF and CSAS expressions have a constant value.

In figure \ref{PlotsSASNum} we plot the ground state energy (at left), the expectation value of the atomic relative population operator (at middle) and the expectation value of the excitation number operator (at right), obtained by numerically minimizing the SAS energy surface in eq. \eqref{EnergySASfin}, all for $N = 2j = 18$ and $k\in \left\lbrace 1,\frac{3}{2},2,\frac{5}{2},3,4,5,7,10 \right\rbrace$. As in the CSAS approach, we use the odd-parity ground state for $k = \frac{3}{2}, \frac{5}{2}$ and the even-parity ground state for $k = 1, 2, 3, 4, 5, 7, 10$. It can be noticed from these plots that the transition looks less smooth, being obvious for the expectation values of the atomic relative population operator and the excitation number operator, as its graphs become discontinuous. The behavior in the normal region is also different from the one shown in figures \ref{PlotsCoherent} and \ref{PlotsSAS}, slightly changing instead of having a constant value. Another interesting thing to observe is the graph of the expectation value of the excitation number operator for $k=1$, as it has a constant value of $\frac{1}{2}$, a completely different behavior than the one shown for all the used approaches and values of $k$.

It is worth mentioning that the cut-off value of the field-matter coupling used in both figure \ref{PlotsSAS} and figure \ref{PlotsSASNum}, is the same as in eq. \eqref{MaximumCoupling}, i.e. the same as in the mean-field case.

\section{\label{QuantumSol} Exact quantum solution}

%%%%%%%%%%%%%%%%%%%%%%%%%%%%%%%%%%%%%%%%%%%%%%%%%%%%%%%%%%%%%%%%%%%%%%%%%%
\begin{figure}[b]
	\includegraphics[width = 8.5cm]{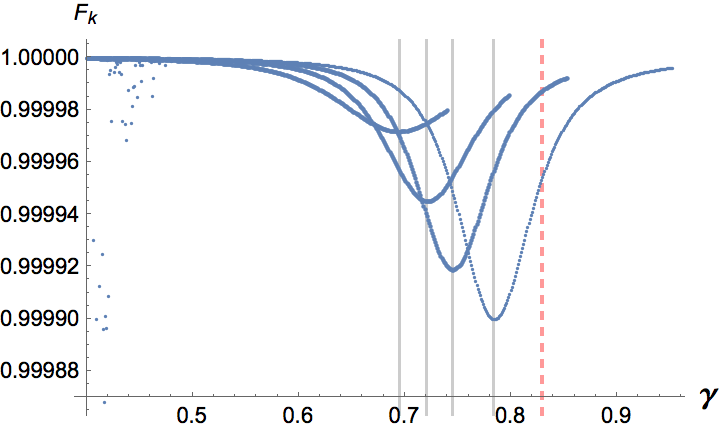}
	\caption{\small Fidelity (blue lines) between neighboring states as a function of the field-matter coupling $\gamma$, obtained using the numerical diagonalization of the Hamiltonian. The vertical lines indicate the transition points characterized by the minimum value of the fidelity, for $k \in \left\lbrace 3,4,5,7\right\rbrace$ from left to right. The vertical red-dashed line corresponds to the transition point in the usual Dicke model, also obtained via numerical means and characterized by the minimum value of its corresponding fidelity.} \label{Fidelidad}
\end{figure}
%%%%%%%%%%%%%%%%%%%%%%%%%%%%%%%%%%%%%%%%%%%%%%%%%%%%%%%%%%%%%%%%%%%%%%%%%%

%%%%%%%%%%%%%%%%%%%%%%%%%%%%%%%%%%%%%%%%%%%%%%%%%%%%%%%%%%%%%%%%%%%%%%%%%%
\begin{figure*}[t]
	a\includegraphics[width = 2.7in]{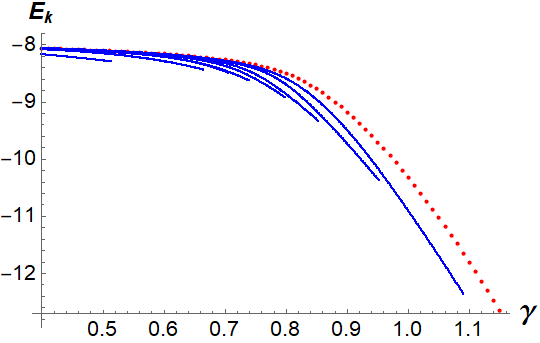} \qquad \qquad \qquad
	b\includegraphics[width = 2.7in]{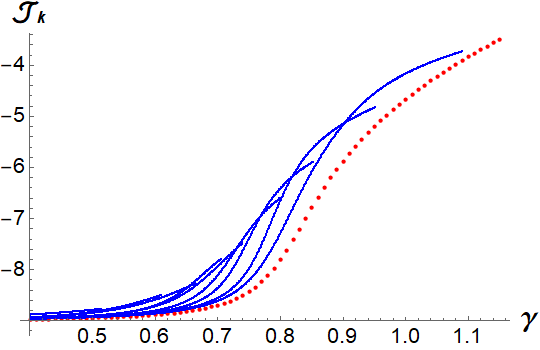}
	c\includegraphics[width = 2.7in]{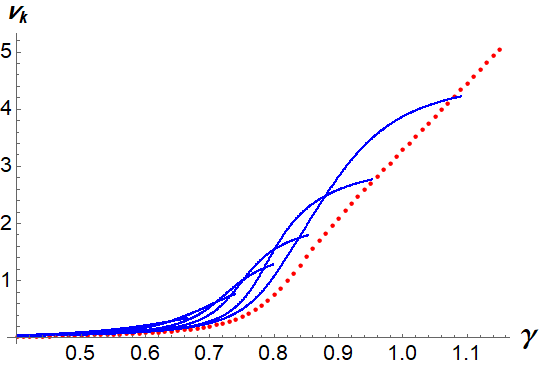}
	\caption{\small Plot of the ground-state energy $E _{k}$ (a), the expectation value of the atomic relative population operator $\mathcal{J} _{k}$ (b) and the expectation value of the analogue of the photon number operator $\nu _{k}$ (c) as a function of the field-matter coupling $\gamma$, obtained by numerically diagonalizing the Hamiltonian, for $k\in \left\lbrace 1,\frac{3}{2},2,\frac{5}{2},3,4,5,7,10 \right\rbrace$ (blue-continuous lines). As a benchmark, the red-dashed lines show these same quantities in the limiting case $k \to \infty$, which corresponds to the HW$(1) \, \otimes \,$SU$(2)$ Dicke model. All plots were obtained using $\omega = 1$, $\Omega = 2$, $N = 18$ and $j = 9$.} \label{PlotsExact}
\end{figure*}
%%%%%%%%%%%%%%%%%%%%%%%%%%%%%%%%%%%%%%%%%%%%%%%%%%%%%%%%%%%%%%%%%%%%%%%%%%

For the exact quantum solution we resort to numerical diagonalization of the Hamiltonian and use the lowest eigenstate to compute the fidelity and the expectation value of the relevant observables.

In figure \ref{Fidelidad} we plot the fidelity between neighboring states as a function of the field-matter coupling parameter $\gamma$, as defined in eq. \eqref{fidelity}. The blue lines correspond to the fidelities for different values of $k$, increasing from left to right, and the gray-continuous vertical lines indicate the corresponding transitions points characterized by the minimum value of the fidelity. The vertical red-dashed line indicates the transition point in the usual Dicke model, also characterized by the minimum value of its corresponding fidelity \cite{Quezada1}. Clearly, the drop in the fidelity is going towards the usual Dicke result as increasing $k$, as it should be.

Notice that the drop in the fidelity is roughly of the order of $10^{-5}$, and it decreases as $k\rightarrow \infty$. However, even in the Dicke model, when a finite number of particles is considered, this drop is relatively small, of the order of $10^{-3}$ for \cite{Bastarrachea} $N=200$. It is only in the thermodynamic limit where the fidelity drops down to zero \cite{Bastarrachea} and both the transition value predicted by the mean-field analysis and the minimum of the fidelity coincide. We expect the $k$DM to have a similar behavior.

Now we analyze how the quantum expectation values of some physical quantities relevant to the system behaves near the transition point. In figure \ref{PlotsExact} we show the exact ground-state energy (at left), the expectation value of the atomic relative population operator $\mathcal{J} _{k}$ (at center) and the photon number operator $\nu _{k}$ (at right), as a function of the field-matter coupling $\gamma$ for a given value of $j$ and different values of $k$. The red dashed lines correspond to the expectation values in the exact quantum solution of the Dicke Hamiltonian. We see from these plots that the quantum solution does indeed change in the normal region, as predicted by the numerically minimized SAS energy surface plots in figure \ref{PlotsSASNum}, eventhough the transition is smoother, which can also be seen from the small ($\approx 10^{-5}$) drop of the fidelity in figure \ref{Fidelidad}.

\begin{figure}[t]
	\includegraphics[width = 8cm]{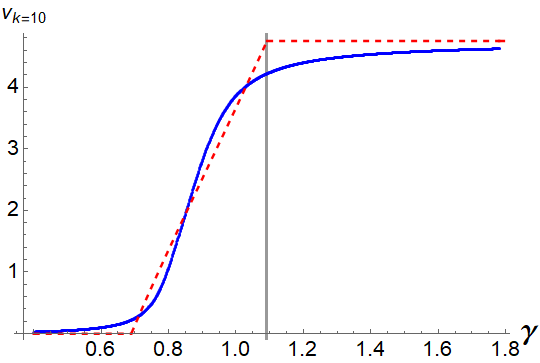}
	\caption{\small Expectation value of the analogue of the photon number operator $\nu _{k}$ as a function of the field-matter coupling $\gamma$, obtained by numerically diagonalizing the Hamiltonian (blue continuous line), and by a mean-field analysis (red-dashed line), both for $k=10$. The vertical line indicates the cut-off value $\gamma_{m}$ of the coupling parameter, according to the mean-field analysis. All plots were obtained using $\omega = 1$, $\Omega = 2$, $N = 18$ and $j = 9$.} \label{comparar}
\end{figure}

It is worth mentioning that the cut-off value of the field-matter coupling parameter $\gamma$ used in the graphs plotted in figure \ref{PlotsExact}, is the one presented in eq. \eqref{MaximumCoupling}, i.e. the same as in the mean-field approach. This causes the blue-continuous graphs to go beyond its permitted values, resulting in crossovers between graphs with different values of $k$ and even between finite-field and Dicke's graphs. The latter allows us to infer that the real cut-off value of the field-matter coupling parameter $\gamma$ is less than the one used in the mean-field approach. However, an explicit calculation of it would require an analytical expression for the real ground state of the system. This, of course, does not mean that the numerical solution after the cut-off value of $\gamma$ is incorrect, as the system is actually solvable for any value of $\gamma$; however, for couplings $\gamma > \gamma_{m}$, the $k$-Dicke model stops showing a super-radiant behavior, even if the system remains in the super-radiant phase, which of course is a direct consequence of having an upper bound in the number of excitations. In figure \ref{comparar} we show a comparison of the expectation value of the analogue of the photon number operator $\nu _{k}$, for $k=10$, between the mean-field and numerical solutions before and after the cut-off value of $\gamma$. In it, the numerical solution shows an asymptotic behavior towards the maximum value attained by the mean-field solution, which we have artificially extend after the cut-off value of $\gamma$, as the mean-field analysis does not allow for $\gamma > \gamma_{m}$.

%%%%%%%%%%%%%%%%%%%%%%%%%%%%%%%%%%%%%%%%%%%%%%%%%%%%%%%%%%%%%%%%%%%%%%%%%%
%%%%%%%%%%%%%%%%%%%%%%%%%%%%%%%%%%%%%%%%%%%%%%%%%%%%%%%%%%%%%%%%%%%%%%%%%%

\section{\label{ConclusionSection} Summary and conclusions}

The use of a fitting of a $k$-oscillator scheme, in conjunction with a given unitary representation of the SU$(2)$ algebra, was proposed to undertake the algebraic study of a nonlinear version of the ubiquitous Dicke model. This group-theoretical extension of the Dicke model, termed the $k$-Dicke model, is based on the fact that the Heisenberg-Weyl algebra HW$(1)$ describing the usual bosonic mode can be obtained by contraction of the SU$(2)$ algebra. The resulting $k$-Dicke Hamiltonian manifests that the dynamical algebra of the extended model is SU$(2) \, \otimes \,$SU$(2)$. Our model resembles the two-fluid Lipkin system of interest in nuclear physics.

It is found that such an algebraic structure, alternative to the quantum algebraic formalism (or to the $f$-deformed algebras), is able to enables us to capture the essence of given nonlinear-oscillator models within a simple algebraic framework, such as those describing a Kerr-type medium and a P\"{o}sch-Teller oscillator. One of the main features of the introduced $k$-oscillator model is that it possesses a maximum number of excitations. This is similar to what occurs with the P\"{o}sch-Teller oscillator, where the maximum number of excitations is identified with the maximum number of bound states. 

We have also discussed how the recent development of new technologies, such as circuit quantum electrodynamics and LC circuits, offer an interesting route for implementing hybrid systems, admiring the possibility of controlling the parameter settings, on demand, and exploring new quantum and semiclassical phenomena. This provides a test bed where many nonlinear quit-oscillator models can be experimentally realized, including the one published in this work.

By using the energy surface minimization method, which consists of minimizing the expectation value of the Hamiltonian in some trial variational state, we found that the $k$-Dicke model exhibits a quantum phase transition at a critical value for the field-matter coupling strength $\gamma _{c} (k)$. The system then transits from the normal phase, for $\gamma$ below the critical value $\gamma _{c} (k)$, to a super-radiant phase, for $\gamma _{c} (k) \! < \! \gamma \! < \! \gamma _{m} (k)$. Here, $\gamma _{m} (k)$ is the maximum value that the field-matter coupling strength can take, and it is a direct consequence of the maximum number of excitations that the $k$-oscillator support. Our analysis was carried out via three methods: through mean-field analysis (i.e. by using SU$(2) \, \otimes \,$SU$(2)$ coherent states), by using parity-preserving symmetry-adapted states (using the critical values obtained in the mean-field analysis and numerically minimizing the energy surface) and by means of the exact quantum solution (i.e. by numerically diagonalizing the Hamiltonian).

We found that all approaches, mean-field, symmetry-adapted and numerical diagonalization, converge to the the Dicke model results as $k\rightarrow \infty$, meaning that in the thermodynamic limit, these different approaches converge to the mean field solution \cite{Quezada1}. From this analysis we can infer that the mean-field quantum phase transition for the $k$-Dicke model remains in the thermodynamic limit.

From figures \ref{PlotsCoherent}, \ref{PlotsSAS}, \ref{PlotsSASNum} and \ref{PlotsExact}, we see that for a finite number of particles, near the transition, all the curves obtained from different approaches look very different. The most reliable approach in this case is of course the numerical diagonalization, nevertheless, it could be a very difficult task since it requires the diagonalization of a $(2j+1) \times (2k+1)$ matrix, with $j=N/2$. On the contrary, both SAS and CSAS are simpler to analyze both numerically and theoretically. In conclusion, for a large number of particles (see Ref. \onlinecite{Bastarrachea}) the expressions obtained in the mean-field analysis provide an excellent description of the behavior of the system near and far from the phase transition. For a small number of particles and specific numerical values of the parameters, if enough computational resources are available, the numerical diagonalization of the hamiltonian is the best choice. If computational resources are limited, then the SAS approach is a better option. Lastly, if analytical expressions are needed, both mean-field and CSAS approaches can be useful, CSAS with the advantage of having a well-defined parity but more complex expressions.

We close by commenting the possible connection between the algebraic model we have considered and the $q$-bosons. The anharmonic $q$-bosons are defined by the following commutation relations \cite{Biedenharn, Macfarlane}
\begin{align}
[a _{q} ,a ^{\dagger} _{q}] = q ^{n _{q}}, \;\;\;\; [n _{q} , a _{q}] = -a _{q} , \;\;\;\; [n _{q}, a _{q} ^{\dagger}] = a ^{\dagger} _{q} , \label{qBosonsAlgebra}
\end{align}
where the deformation parameter $q$ is in general a complex number. This algebra, which is usually referred as Heisenberg-Weyl $q$-algebra and denoted by HW$_{q}$, defines a special deformation scheme of the usual harmonic-oscillator algebra (\ref{HW(1)}), to which it is reduced in the limit $q \to 1$. It has been useful in the description of anharmonic vibrations of diatomic molecules and solids, since the corresponding oscillator-like model exhibits the main features of the Morse oscillator. Our $k$-oscillator algebra (\ref{b-algebra}) may be obtained from the $q$-bosons (\ref{qBosonsAlgebra}) for real values of $q$ approaching to one from below. To see this, let $q<1$ and $p = \frac{1}{1-q}$, so that $q = 1 - \frac{1}{p}$. Then we have
\begin{align}
q ^{n _{q}} = \left( 1 - \frac{1}{p} \right) ^{n _{q}} . \label{App1}
\end{align}
The harmonic limit is recovered for $p \to \infty$ in this parametrization. Further, assuming that $(1/p) \ll 1$, we can Taylor expand the above result to obtain
\begin{align}
q ^{n _{q}} \approx 1 - \frac{n _{q}}{p} + \frac{1}{2} \frac{n _{q} (n _{q} - 1)}{p ^{2}} + \mathcal{O} (p ^{-3}) . \label{App2}
\end{align}
If we now substitute the leading terms (up to order $1/p$) in the $q$-bosons algebra (\ref{qBosonsAlgebra}) and we identify the parameter $p$ with $k$, $n _{q}$ with $n _{k}$ and the creation and annihilation operators $a _{q} ,a ^{\dagger} _{q}$ with $b _{k} , b ^{\dagger} _{k}$, we recover the SU$(2)$ algebra (\ref{b-algebra}). We mention in passing that, if $q$ approaches to one from above, we can define the new parameter $r = \frac{1}{q-1}$ and the approximation equivalent to eqs. \eqref{App1} and (\ref{App2}) leads to the non-compact SU$(1,1)$ algebra. In summary:
\begin{align}
\lim _{q \to 1 - 0 ^{+}} \mbox{HW} _{q} \to  \mbox{SU}(2) , \quad \lim _{q \to 1 + 0 ^{+}} \mbox{HW} _{q} \to  \mbox{SU}(1,1) .
\end{align}
Although the $q$-bosons algebra contracts to the SU$(2)$ algebra, it is not clear that their representations also match in the limit. This can be seen from the fact that the SU$(2)$ algebra is naturally defined in a finite Hilbert space, while the $q$-algebra can be defined in a finite-dimensional representation only when $q$ is a nontrivial root of the equation \cite{Angelova} $q ^{n _{q}} = 1$. A possible extension of this work must be then the analysis of the quantum phase transition in the $q$-Dicke model.

\begin{acknowledgments}
L. F. Q. thanks C3-UNAM for financial support. A. M. R. acknowledges support from DGAPA-UNAM project No. IA101320. We thank R. Le\'{o}n-Montiel for useful discussions.
\end{acknowledgments}

%\section*{AIP Publishing Data Sharing Policy}
%Data sharing is not applicable to this article as no new data were created or analyzed in this study.\\

\end{document}